\documentclass[%
 preprint,%
 amssymb, amsmath,%
 prl,cha,superscriptaddress,%
]{revtex4-1}
\usepackage{graphicx}
\usepackage{comment}
\usepackage[capbesideposition=right]{floatrow}
\usepackage{sidecap}
\usepackage{nicefrac}
\usepackage{docs}%
\usepackage{bm}%
\usepackage[colorlinks=true,linkcolor=blue]{hyperref}%
\expandafter\ifx\csname package@font\endcsname\relax\else
 \expandafter\expandafter
 \expandafter\usepackage
 \expandafter\expandafter
 \expandafter{\csname package@font\endcsname}%
\fi
\usepackage{color}

\DeclareOldFontCommand{\bf}{\normalfont\bfseries}{\mathbf}
\DeclareOldFontCommand{\rm}{\normalfont\rmfamily}{\mathrm}

\usepackage{stackengine}
\def\cpm{\mathbin{\ensurestackMath{\abovebaseline[-1.9pt]{%
\stackunder[-1.5pt]{\color{red}+}{\color{black}-}}}}}

\begin{document}
	
\title{Scaling of intrinsic domain wall magneto-resistance with confinement in electromigrated nanocontacts}%



\author{Robert M. Reeve}%

\affiliation{Institut f\"ur Physik, Johannes Gutenberg-Universit\"at Mainz, 55099 Mainz, Germany}%

\affiliation{Graduate School of Excellence Materials Science in Mainz, 55128 Mainz, Germany}%

\author{Andr\'{e} Loescher}%

\affiliation{Institut f\"ur Physik, Johannes Gutenberg-Universit\"at Mainz, 55099 Mainz, Germany}%

\author{Hamidreza Kazemi}%

\affiliation{Physics Department and Research Center OPTIMAS, University of Kaiserslautern, 67663 Kaiserslautern, Germany}%

\author{Bertrand Dup\'e}%

\affiliation{Institut f\"ur Physik, Johannes Gutenberg-Universit\"at Mainz, 55099 Mainz, Germany}%
\affiliation{Graduate School of Excellence Materials Science in Mainz, 55128 Mainz, Germany}%

\author{Thomas Winkler}%

\affiliation{Institut f\"ur Physik, Johannes Gutenberg-Universit\"at Mainz, 55099 Mainz, Germany}%

\author{Daniel Sch\"{o}nke}%

\affiliation{Institut f\"ur Physik, Johannes Gutenberg-Universit\"at Mainz, 55099 Mainz, Germany}%

\author{Jun Miao}%

\affiliation{School of Materials Science and Engineering, University of Science and Technology Beijing, Beijing 100083, China}%

\author{Kai Litzius}%

\affiliation{Institut f\"ur Physik, Johannes Gutenberg-Universit\"at Mainz, 55099 Mainz, Germany}%

\affiliation{Graduate School of Excellence Materials Science in Mainz, 55128 Mainz, Germany}%

\author{Nicholas Sedlmayr}%

\affiliation{Department of Physics and Medical Engineering, Rzesz\'ow University of Technology, 35-959 Rzesz\'ow, Poland}%

\author{Imke Schneider}%

\affiliation{Physics Department and Research Center OPTIMAS, University of Kaiserslautern, 67663 Kaiserslautern, Germany}%

\author{Jairo Sinova}%

\affiliation{Institut f\"ur Physik, Johannes Gutenberg-Universit\"at Mainz, 55099 Mainz, Germany}%

\affiliation{Graduate School of Excellence Materials Science in Mainz, 55128 Mainz, Germany}%

\author{Sebastian Eggert}%

\affiliation{Physics Department and Research Center OPTIMAS, University of Kaiserslautern, 67663 Kaiserslautern, Germany}%

\author{Mathias Kl\"aui}
\email{klaeui@uni-mainz.de}
\affiliation{Institut f\"ur Physik, Johannes Gutenberg-Universit\"at Mainz, 55099 Mainz, Germany}%
\affiliation{Graduate School of Excellence Materials Science in Mainz, 55128 Mainz, Germany}%

\date{2018}%

\begin{abstract}
In this work we study the evolution of intrinsic domain wall magnetoresistance (DWMR) with domain wall confinement. Clean permalloy notched half-ring nanocontacts are fabricated using a special ultra-high vacuum electromigration procedure to tailor the size of the wire in-situ and through the resulting domain wall confinement we tailor the domain wall width from a few tens of nm down to a few nm. Through measurements of the dependence of the resistance with respect to the applied field direction we extract the contribution of a single domain wall to the MR of the device, as a function of the domain wall width in the confining potential at the notch. In this size range, an intrinsic positive MR is found, which dominates over anisotropic MR, as confirmed by comparison to micromagnetic simulations. Moreover, the MR is found to scale monotonically with the size of the domain wall, $\delta_{DW}$, as 1/$\delta_{DW}^b$, with $b=2.31\pm 0.39 $. The experimental result is supported by quantum-mechanical transport simulations based on ab-initio density functional theory calculations.

\end{abstract}
\maketitle
\section{Introduction}

Magnetoresistance (MR) effects encompass a range of varied phenomena which are of great fundamental interest, but also of significant practical importance for detecting magnetic states in devices. Prominent examples of effects that are employed in current technology include giant magnetoresistance (GMR)~\cite{GMR1, GMR2}, tunneling magnetoresistance (TMR)~\cite{TMR1, TMR2} and anisotropic magnetoresistance (AMR)~\cite{AMR}. Over the last years, a number of devices have been proposed relying on magnetic domain wall (DW) propagation in nanowires~\cite{racetrack, DWlogic}, making domain walls interesting elements for achieving different functionalities. Domain walls have associated MR contributions which can be employed for sensing a particular state or position of a DW in a wire and for memory functionality~\cite{DWMRmemory, MK}. A wide variety of theoretical models exist to explain such DWMR, predicting different sizes and even signs of the MR depending on the materials system and the dimensions of a given device~\cite{CabFal, Viretmistracking, LevyZhang,negposDWMRbandBauer, negposDWMRbandBelzig, negDWMRweaklocal, banddDWMR} as reviewed in Refs.~\cite{MarrowsDWreview, KentRev}. In general, larger effects are predicted and observed as the confinement of a DW is increased on reducing the size of a wire or for domain walls at constrictions~\cite{DWMRnano, nanoconstrictMR, BalDWMRscaling, AMRnanocontactsKlaui}. This includes enhanced AMR effects~\cite{nanoAMR, GAMRatomic, PointAMR} and new ballistic magnetoresistance contributions (BMR)~\cite{Qspinvalve, ViretfirstBMR}, providing an avenue for tailoring DWMR which exhibits favourable scaling behaviour on reducing the size of the structures, as desired for miniaturized devices. However in particular for small nanocontacts, it is often difficult to disentangle other contributions to the resistance which can manifest as apparent MR effects, such as mechanical rearrangements of a system~\cite{BMRQ,magnetostrictionswitching, NoNiBMR, BMRartifacts, microspheres}. Furthermore the influence of contamination or local oxidation can often dominate the signals~\cite{MoistureEM,gasatomicquantization, DWMRoxy, BMRoxy}, making an understanding of the fundamental intrinsic properties of a system difficult and requiring careful experimental design in order to exclude or account for such parasitic contributions. As a result, and due to the large range of approaches to determine MR in confined magnetic systems, there is a large spread in the reported values~\cite{negDWMRexp, negDWMRexpCoZigzag, NegDWMRParkin, negDWMRexpOno, Viretmistracking,posDWMRexpCo, posandnegDWMRexpCo, negandposDWMRexpFe, largeBMRexp, NiBMRexp, smallerBMR, KentRev, negDWMRpy, AMRnanocontactsKlaui} and hence it can be difficult to understand the regimes of applicability of different theories. One proven approach to a robust determination of different contributions to the MR involves the formation of magnetic Py nanocontacts in ultra-high vacuum (UHV), with different regimes of behaviour observed depending on the contact size. For larger contacts no significant intrinsic DWMR is observed with only relatively small effects, which are of less interest for devices, and which can be accounted for due to the AMR effect from the components of the magnetization within the wall which are no-longer collinear with the current~\cite{AMRnanocontactsKlaui, fieldMEdepinKlaeui,negDWMRexpCoZigzag}. Alternatively for ultra-narrow contacts much larger intrinsic effects have been reported~\cite{BMRKlaui}, yet these effects are very sensitive to the exact atomic-coordination at the domain wall position and change in both magnitude and sign based on small atomic rearrangements~\cite{AchillesNi}, hence they are difficult to employ in a device context. What is not yet clear is the transition between these two regimes of behaviour in this system and the structure widths and resulting resistance values at which DWMR starts to dominate. In particular, certain theories predict a regime where the DWMR should scale monotonically with the size of the domain wall~\cite{Viretmistracking, LevyZhang, banddDWMR, wireDWMR, bandsDWMR, bandsDWMR2} which would be highly advantageous, yet while isolated reports of large DWMR for narrow DWs exist, the different materials systems and geometries employed make it hard to draw firm conclusions. Surprisingly, there are few studies systematically investigating the scaling behaviour of the DWMR with DW size, which is a key component of the theories. For nanowires with perpendicular magnetic anisotropy (PMA), the scaling of DWMR with DW width has been explored by tailoring the domain wall size via ion-irradiation induced anisotropy modulation in a multilayer system~\cite{TuneDWMR}. However for in-plane systems and the alternative approach of tailoring the domain wall size via domain wall confinement in narrow wires or in notches~\cite{Bruno, SEMPADWsize}, the scaling relation remains to be robustly demonstrated. Yet this approach of DW tailoring is particularly compatible with various proposed device architectures~\cite{notch1, racetrack}. 

In this work we employ our previously established approach of tailoring the size of magnetic nanocontacts in clean UHV conditions~\cite{NanoReeve, AMRnanocontactsKlaui, BMRKlaui} to study the evolution of intrinsic DWMR for nanocontacts in a regime where the contacts are expected to be a few nm in size. The resulting contacts are smaller than those that can be achieved by direct lithographic patterning techniques, but not so small that tunneling or atomic coordination effects dominate the signals. We compare the resistance of the samples with and without a magnetic domain wall at the contact position where, as a result of the geometrical confinement, the domain wall width is similarly tailored, as confirmed by micromagnetic simulations. The results reveal an unambiguous contribution from intrinsic DWMR and a monotonic evolution in the MR with the DW size. We discuss the origin of these effects in relation to the results of micromagnetic simulations and compare the results to the predictions of quantum-mechanical scattering calculations in order to understand the regimes of validity of the experimentally observed contributions.

%

\section{Methods}
\label{sec:exp}

\subsection{Experiments}

\begin{figure}[tb!]
\centering
\includegraphics[width=0.9\textwidth,keepaspectratio=true]{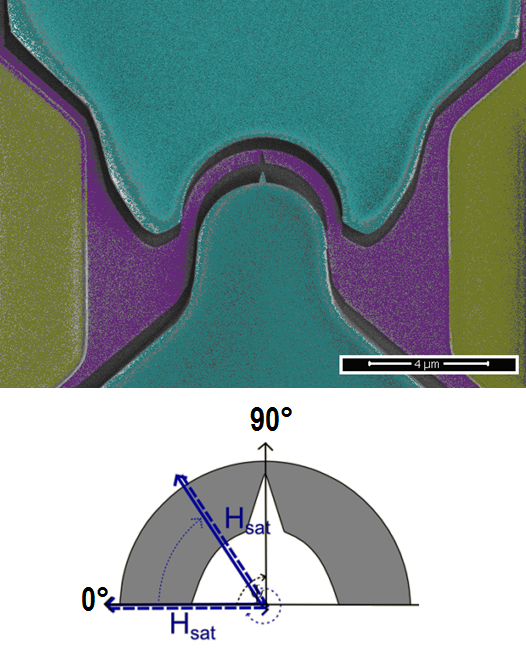}
\caption{False colour scanning electron microscope image of a typical structure. The sample consists of a notched Permalloy half-ring of radius 2.5 $\mu$m and 400 nm width with initial constriction size $\sim$70$\,$nm (purple). Due to the large undercut of the resist, the sample is electrically isolated from the permalloy on top of the resist (green), enabling direct in-situ measurement. Electrical contact is made via  Cr(5 nm)/Au(55 nm) contact pads (yellow). The schematic below demonstrates the \emph{mode-\'{e}toile} measurement scheme, as described in the text.}
\label{SEM}
\end{figure}

The experiments were carried out on polycrystalline permalloy (Py: Ni$_{80}$Fe$_{20}$) samples of 23$\,$nm thickness, grown by thermal evaporation in a UHV chamber with a base pressure of $5\times10^{-10}\,$mbar. In order to avoid certain parasitic contributions to the MR measurements, we use our previously established procedure to fabricate magnetic nanocontacts in clean UHV conditions ~\cite{BMRKlaui} and perform in-situ magnetic characterization, as outlined below. The size of the contact is subsequently tailored in-situ from the initial lithographically defined cross-section, which is a few tens of nm at the narrowest part, down to the single nm regime and below via a computer-regulated electromigration procedure~\cite{initialcontactEM, TempContEM, goldEM, SPMnanoEM}. At each stage, the MR behaviour is measured to determine the evolution of the properties with contact size. In order to avoid magnetostrictive contributions we choose Py as our magnetic material and use a notched half-ring geometry where we can control the presence and position of DWs in the structure at remanence, as depicted in Figure~\ref{SEM}. The system and process of nanocontact fabrication are described in detail in~\cite{NanoReeve}. In the initial state, the contact is around  $70\;$nm wide at the notch position. Both electromigration and magnetotransport measurements are performed at temperatures around 80$\;$K using a liquid nitrogen cryostat. In order to characterize the magnetoresitance we perform a so-called \emph{mode-\'{e}toile} measurement~\cite{Pfeiffer}.  A magnetic field of 150$\;$mT is applied along a given angle, $\theta$, using a vector magnet~\cite{grain} and then relaxed to zero, before measuring the two-point resistance of the contact at remanence. This process is then repeated for incrementally changing angles until the desired angular range is probed. The electromigration procedure is subsequently performed in order to narrow down the sample at the region of highest current density, which corresponds to the notch, as detected by the concurrent increase in resistance of the sample from the initial state of around 436~$\Omega$. The magnetotransport characterization is then carried out for the new contact state and the whole process is iterated until the contact size has been reduced to the expected sub-nm regime~\cite{NanoReeve}.

\subsection{Density function theory calculations (DFT)}

To theoretically understand the transport properties of the system semi-quantitatively, we have modeled Ni$_{80}$Fe$_{20}$ by Ni$_3$Fe in the common faced-centered crystal structure (fcc). In the experiment, the transport at the narrow-constriction is likely to be dominated by a single grain~\cite{grain}, while the 5$\%$ composition difference is not expected to give a significance change in the transport properties. In this configuration, the Fe atoms sit at the summit of the pseudo-cubic unit cell and the Ni atoms are at the center of the faces. The experimental pseudo-cubic lattice parameter $a= 3.55$\,\AA\,was used. We have calculated the band structure with the full potential linearized augmented plane wave (FLAPW) basis set as implemented in the FLEUR~\footnote{\lowercase{www.flapw.de}} ab-initio package. We have used a cutoff of the plane wave basis set of $K_{\mathrm{max}}=5.4$\,bohr${}^{-1}$ and 560 k-points in the irreducible Brillouin zone. The band structure was obtained with the Perdew-Burke exchange and correlation~\cite{Perdew1992b}.

\begin{figure}[t]
\centering
\includegraphics[width=1.0\textwidth]{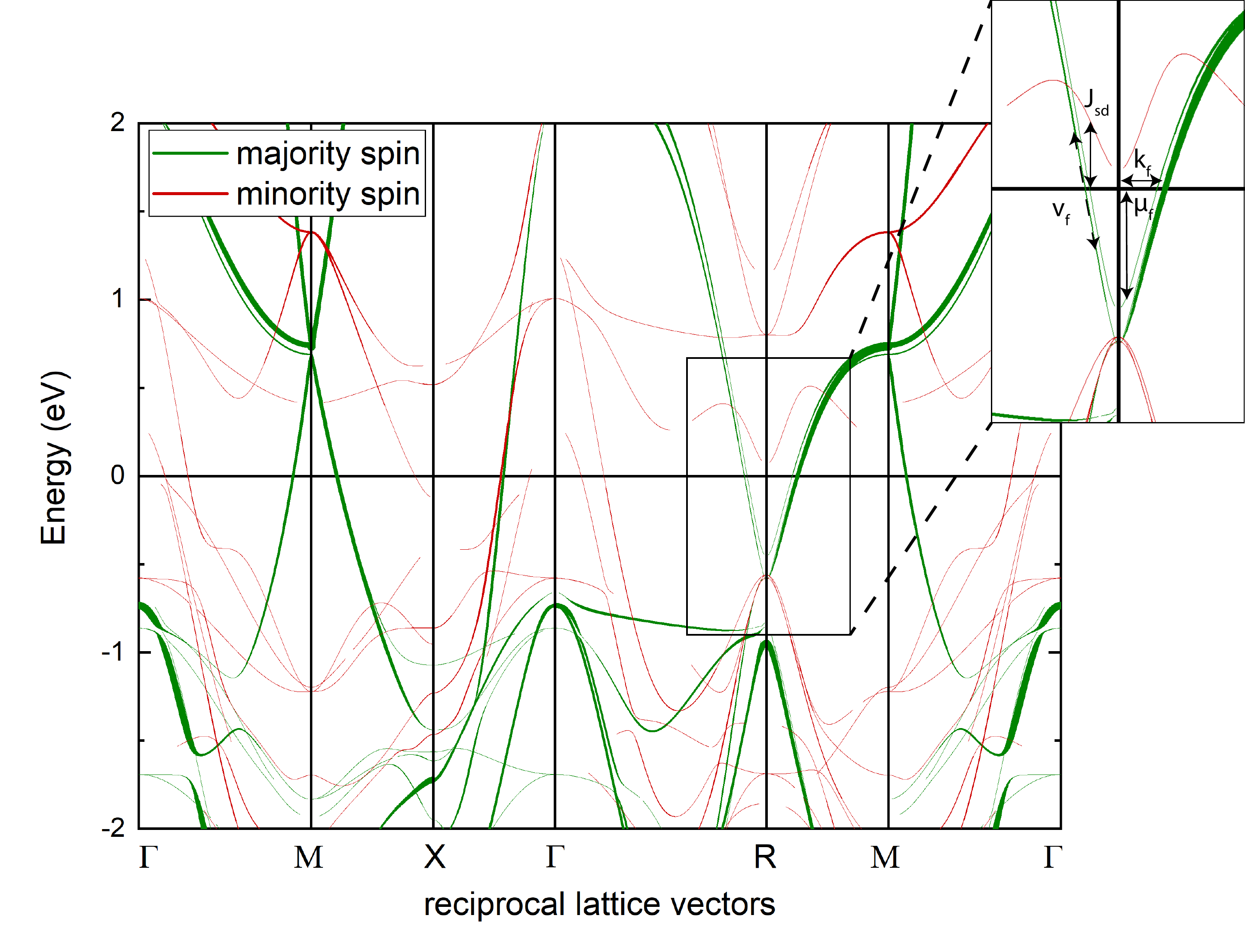}
\caption{\label{fig:band:struct} Band structure of Ni$_3$Fe. The width of the lines corresponds to the s-character of the Bloch states. The inset corresponds to the region of the band structure that was used to extract the Fermi velocity $v_{\mathrm{f}}$, the \emph{sd} coupling $J_{\mathrm{sd}}$, the chemical potential $\mu$ and the Fermi wavevector $k_{\mathrm{f}}$.}
\end{figure}

The band structure was extrapolated using maximally localized Wannier functions as implemented in the wannier90 package~\cite{Marzari1997,Souza2001,Mostofi2014}. We have used 9 Wannier orbitals (1s, 3p and 5d) for each atom in the chemical unit cell for spin up and spin down. The bands were disentangled within an energy window up to 30 eV containing 50 bands and a frozen window up to 15 eV above the Fermi level~\cite{Souza2001}. To obtain the \emph{s}-character contribution of the Bloch state shown in Figure~\ref{fig:band:struct}, we have projected the resulting wannier orbitals on the \emph{s}-wannier orbitals.

Figure~\ref{fig:band:struct} shows the \emph{s}-character of the Bloch states along the high symmetry lines of a simple cubic lattice. The conduction bands are evenly spread along the high symmetry lines. To parametrize a simple tight-binding Hamiltonian, it is convenient to pinpoint bands which have the same orbital character and which are not degenerate at the Fermi level. The bands around the $R$-point show these characteristics (Fig.~\ref{fig:band:struct} inset). We can then extract the Fermi velocity,  $\bm{v}_{\mathrm{f}}=\tfrac{\mathrm{d} \epsilon_{nk}}{\hbar \mathrm{d} \mathbf{k}}$, the s-d coupling, $J_{\mathrm{sd}}$ which corresponds to the band splitting at the Fermi level and the chemical potential $\mu$.
We have focused on the up and down spin channels at the Fermi level along the high symmetry line $\Gamma-R$. We have extracted  $v_{\mathrm{f}}=0.93 \times 10^6$\,ms$^{-1}$ which is of the same order of magnitude as typical metals~\cite{Ashcroft1976}, $J_{\mathrm{sd}}=0.29$\,eV, $k_{\mathrm{f}}=0.084$\,bohr${^{-1}}$ and $\mu=0.45$\,eV. These parameters were then used to parametrize an effective tight-binding Hamiltonian.

\subsection{DC electrical conductivity from quantum simulations}
\label{sec:DcConductivity}

To obtain insights into the domain wall magnetoresistance (DWMR), we perform transport calculations by solving the quantum-mechanical scattering in the ballistic regime for both spin channels through a domain wall. The ballistic regime approximation is justified by the spin diffusion length on the order of $0.6-14$\,nm~\cite{basspratt,mfpref2} which has been measured in Ni$_{80}$Fe$_{20}$ and compares favourably with the experimentally estimated domain-wall sizes down to $\sim1$\,nm (see Ref.~\cite{NanoReeve} and Fig.~\ref{mainplot}).

According to our DFT calculations the physical behaviour of low energies can be described by a surprisingly simple Hamiltonian, which is commonly referred to as the s-d model~\cite{zhang2004roles}. In this approximation the $d$-orbital is associated to the magnetic moments of permalloy which have a fixed position and direction, while the spin of the $s$-electrons is scattered through the domain wall. In the case of non-collinear magnetization, such a model can be written as:

\begin{eqnarray}\label{eq:Hamiltonian}
	\hat{H} & = & 
	\sum_{j,\sigma} \left[ -t \left( \hat{c}^{\dagger}_{j\sigma} \hat{c}_{(j+1)\sigma} + \textrm{h.c.} \right) - \mu  \hat{c}^{\dagger}_{j\sigma} \hat{c}_{j\sigma} \right]\\
	& & -J_{\rm sd} \sum_{j,\sigma,\sigma'} \left( \hat{c}^{\dagger}_{j\sigma} \boldsymbol{S}_{\sigma\sigma'} \cdot \boldsymbol{n}_j \hat{c}_{j\sigma'} \right)\nonumber,
\end{eqnarray}
where $t$ is the hopping parameter, $\mu$ is the chemical potential, $J_{\mathrm{sd}}$ is the spin-splitting, $\boldsymbol{n}_j$ is a unit vector $\boldsymbol{M}_j/M_s$ at site $j$ and $\boldsymbol{S}_{\sigma\sigma'}$ is the spinor of the $s$-electron. To derive the suitable scattering transport parameters, we consider the band-structure of Ni$_{3}$Fe, which was obtained by DFT calculations. The value of the $J_{\mathrm{sd}}=0.29$\,eV and $\mu=0.45$\,eV can be used directly but the $v_{\mathrm{f}}=0.91 \times 10^6$\,ms${^{-1}}$ and $k_{\mathrm{f}}=0.084$\,bohr${^{-1}}$ have to be transformed to a hopping parameter $t$. For a free electron, the dispersion $ \epsilon \left( k \right) $ is given by the simple relation:

\begin{equation}
\epsilon = -2t \cos{k^{\rm l}_{\uparrow} a} - \frac{J_{\rm sd}}{2} - \mu 
\end{equation}
where $a=3.55$\,\AA~is the lattice parameter of bulk Ni$_3$Fe. This relation can be used to obtain the Fermi velocity $\bm{v}_{\mathrm{f}}=\tfrac{\mathrm{d} \epsilon}{\hbar \mathrm{d} \mathbf{k}}$ as a function of $k_{\mathrm{f}}$. We therefore obtain
\begin{equation}
t= \frac{ \hbar v_{\mathrm{f}} }{ 2 a \sin \left( k_{\mathrm{f}} a \right) }
\end{equation}
This yields $t=1.58$ eV. The DFT calculations were done for a collinear magnetic texture. We impose the DW profile within the Hamiltonian from eq.~\ref{eq:Hamiltonian} via the $J_{\mathrm{sd}}$-coupling. Without loss of generality, the magnetization is constrained in the $ \left( x ; z \right) $ plane and is given by:
\begin{equation}
\boldsymbol{n}_j=\boldsymbol{M}_j/M_{\rm s} \equiv \sin(\Theta(z))\hat{\bf z}+\cos(\Theta(z))\hat{\bf x}
\end{equation}
where $\Theta(x)$ is the polar angle that changes from $0$ to $\pi$ over the DW's width $\delta_{DW}$ or a N\'eel DW.

In order to find the total resistance we have to average over the resistance $R(\vec{k}^{\rm l}_\sigma)$ for different incident particles with momentum $\vec{k}^{\rm l}_\sigma$ over the 3D Fermi sphere~\cite{Ashcroft1976,stiles2002anatomy},
\begin{equation}\label{eq:theoDWMR}
	\frac{1}{R_{\rm DW}}=e^2 \int \left[ \frac{d \vec{k}^{\rm l}_\sigma}{8\pi} \frac{1}{R(k^{\rm l}_\sigma)} \nu(\vec{k}^{\rm l}_\sigma)\left( \frac{\partial f(\vec{k}^{\rm l}_\sigma)}{\partial \vec{k}^{\rm l}_\sigma} \right) \right],
\end{equation}
where $e$ is the electron charge, $\nu(\vec{k})$ is the velocity and $f(\vec{k})$ is the Fermi-Dirac distribution. In our case, the scattering parameters only depend on the $\vec{k}_\sigma$-components along the domain wall if the perpendicular components are approximately conserved.  It is therefore sufficient to solve the 1D scattering problem using the usual plane-wave ansatz: 
\begin{equation}\label{eq:ansatzplainwave}
	\begin{cases}
			\psi_{j\sigma}^{\rm l} = a_{\sigma} e^{ik_{\sigma}^{\rm l}j} + r_{\sigma} e^{-ik_{\sigma}^{\rm l}j}, & j \leq 0, \\
			\psi_{j\sigma}^{\rm r} = t_{\sigma} e^{ik_{\sigma}^{\rm r}j}, & j \geq \delta_{DW} ,
	\end{cases}
\end{equation}
where the wave number $k_\sigma = \hat{\bf z} \cdot \vec{k}^{\rm l}_\sigma$ is the projection of $\vec{k}_\sigma$ along the domain wall direction and we consider a completely up-spin-polarized current $a_\uparrow=1 , a_\downarrow=0$. Note that the reflection and transmission coefficients are also a function of incident momentum.  Here the wave-numbers for up and down spin must be related by the dispersion relations

\begin{align} \label{eq:dispersion}
	\epsilon + \mu &= -2t \cos{k_{{\color{red}\downarrow}\uparrow a}^{\rm l}}\cpm \frac{J_{\rm sd}}{2},\\
	\epsilon + \mu &= -2t \cos{k_{{\color{red}\uparrow}\downarrow a}^{\rm r}}\cpm \frac{J_{\rm sd}}{2}.
\end{align}

Finally, knowing the resistance is directly proportional to the reflection coefficients, the resistance for each momentum can be determined from,
\begin{equation}
	R(k^{\rm l}_\sigma) \propto \frac{\sum_\sigma |r_{\sigma} |^2\sin(k^{\rm l}_{\sigma})}{\sum_\sigma |a_{\sigma} |^2\sin(k^{\rm l}_{\sigma})}.
\end{equation}
Since the scattering problem is solved for a number of incident particles with different momenta numerically, the integral is turned into a summation over the values obtained for a reasonably large number of incident particles (here we found out that 150 different $k$-values suffice). From Eq. \ref{eq:theoDWMR} and after going to  polar coordinates, the resistance for each $\delta_{\rm DW}$ averaged over different momenta is simplified as,
\begin{equation}
    \frac{1}{R_{\rm DW}} = \int_0^1 {\rm d}z k_{\rm f}^2 \frac{v(k_{\rm f} z)^2}{R(k_{\rm f} z)}
\end{equation}\label{eq:totalDWMR}where $z=\cos{\theta}$ parameterizes the incoming direction and we have approximated $\frac{\partial f(\vec{k}^{\rm l}_\sigma)}{\partial \vec{k}^{\rm l}_\sigma}$ with a delta function which is justified for the relevant parameter ranges. The results and their comparison to experiments are discussed in Sec. \ref{sec:dis}. 

\subsection{Micromagnetic Simulations}
Micromagnetic simulations of the domain wall spin structure were performed using the multi-scale modified MicroMagnum code~\cite{MM} as described in Ref.~\cite{Andrea}. Multiscale modelling is necessary to realistically model the spin structures even for very narrow domain walls. Standard parameters for permalloy were chosen, namely exchange energy constant, $A=13\times10^{-12}$\,J/m$^{3}$, anisotropy constant, $K=0$ and saturation magnetization, $M_s=8\times 10^5$A/m. In-plane cell sizes of $2\times2$\,nm  and $0.4\times0.4$\,nm were used for the coarse- and fine- scale regions, respectively, with the cell size normal to the plane equal to the film thickness. A high damping, $\alpha=0.5$, was used to speed up the convergence. In order to further save on simulation time we simulated just the central portion of the half-ring over an area of 2000$\times$720\,nm and fixed the magnetization at both ends of the wire to be parallel to the wire axis, as expected from the shape anisotropy and confirmed previously by imaging of the spin structures. In the simulations this was achieved by applying a localized large field to the cells at the two edges. The fine-scale region of the simulation is applied to a  region of 40$\times$40\, nm for the portion of the wire where the confined DW is expected, based on initial coarse-scale simulations of the whole system. For most constriction widths this corresponds to the central notch region, however for the asymmetric transverse wall in the case of the smallest constriction, the fine-scale region was positioned on the outer-edge of the half-ring. To account for the effect of electromigration on the dimensions of the contact, as employed in the experiments to change the contact size, we reduce the width of the wire at the notch position and since the electromigration procedure is expected to reduce the size of the contact in all dimensions, for widths smaller than the initial wire thickness we also reduce the simulated wire thickness accordingly. To calculate the AMR contribution from the simulated magnetization configurations for states with and without a DW at the notch we perform simulations of the current distribution in the samples. From these, we calculate the difference in the resistance in each case, taking into account the angle between the magnetization and the current and assuming an AMR of around 1\%, as in Ref. \cite{AMRcalc}.

\section{Results}
\label{sec:res}

\begin{figure}[tb!]
\centering
\includegraphics[width=0.7\textwidth,keepaspectratio=true]{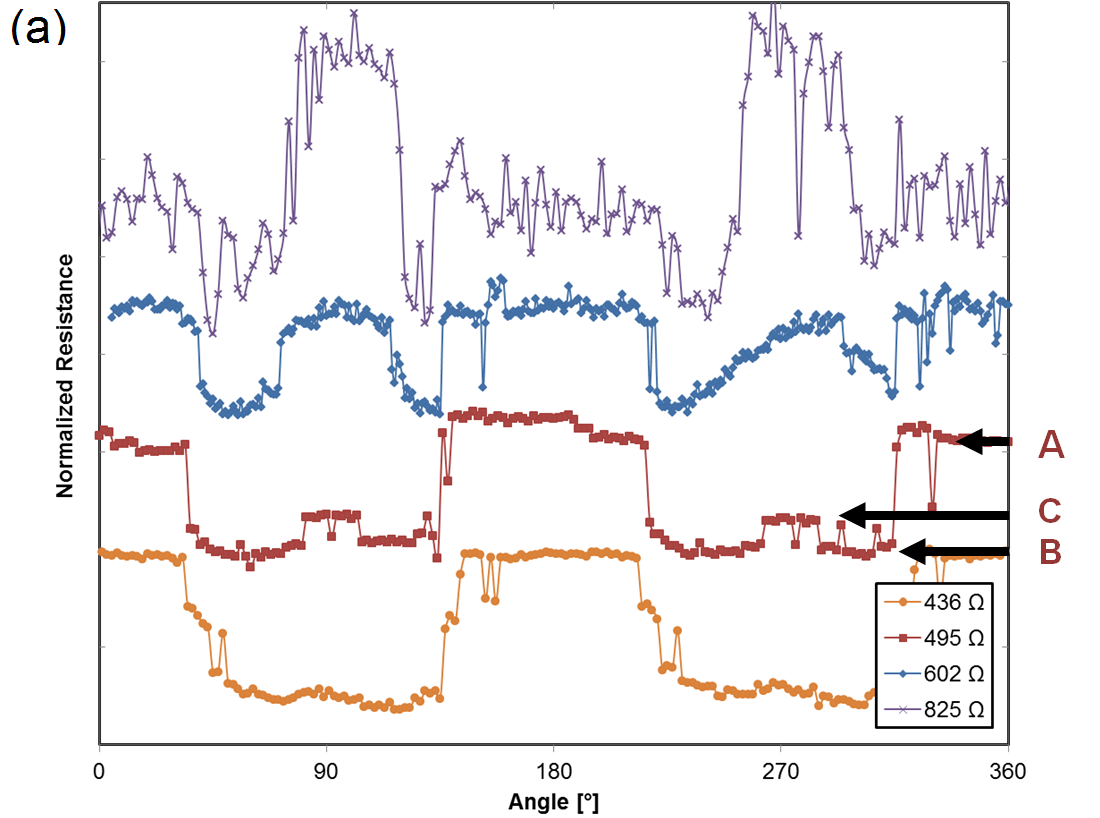}
\includegraphics[width=0.7\textwidth,keepaspectratio=true]{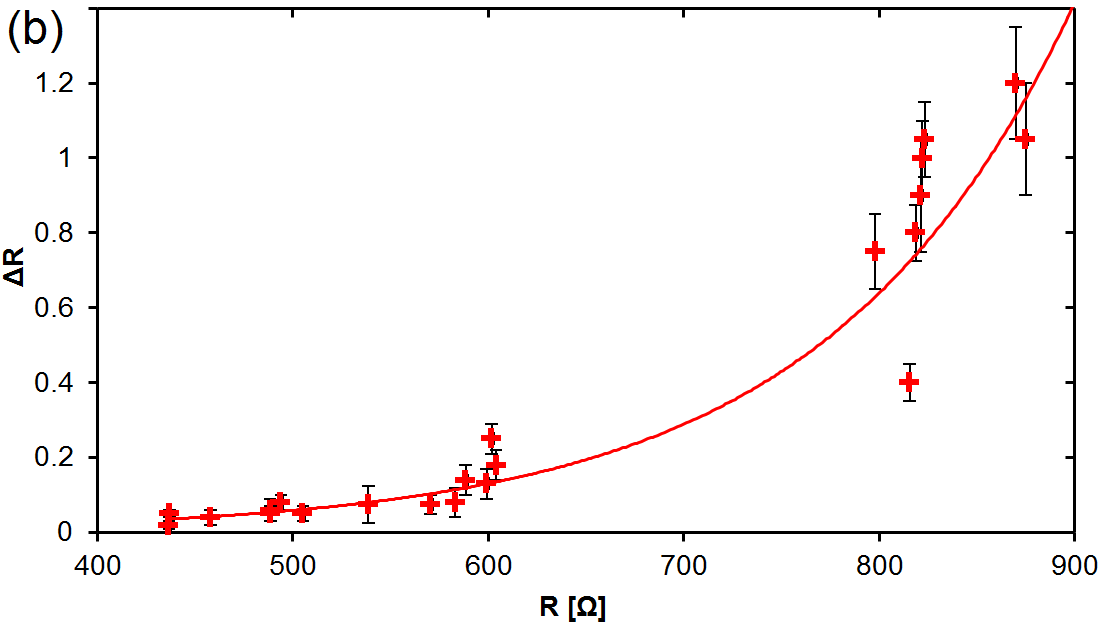}
\caption{(a) Normalized resistance of the contact at remanance as a function of angle following saturation along the given angle. Four different base resistance levels are shown, corresponding to different resistance states for contacts without a domain wall, equivalent to 436, 496, 602 and 825 $\Omega$ from bottom to top. The plots are offset for clarity. Three main levels are seen, depending on the state, as indicated for the 495 $\Omega$ resistance sample. A) A high resistance state when there is no DW in the structure. B) A lower resistance state when a DW has been nucleated in the wire, away from the notch region. C) A variable sized local-peak corresponding to the DW present in the notch. A continuous thermal drift in resistance as a result of heating from the magnet has been subtracted from the presented data in each case, from a fit to the resistance levels in the regions without a DW in the structure.
(b) Extra resistance for a DW in the notch compared to a DW in the wire ($\Delta R=(C-B)/A$) as a function of the contact resistance, $R$. The line is a guide to the eye.}
\label{etoile}
\end{figure}

The results of the \emph{mode-\'{e}toile} measurement for selective resistance states of the sample are presented in Figure~\ref{etoile} (a)~\cite{Loescherthesis}. For the initial state of the contact - i.e. the state with a resistance of 436 $\Omega$, the plot shows two main resistance levels as a function of the angle, which can be understood as follows: Depending on the angle of the saturating field,  a domain wall may or may not be nucleated in the structure and as a result the two resistance states can be attributed to the presence and absence of a domain wall between the probes~\cite{AMRnanocontactsKlaui}. When the field is initially applied along angles in the range ~325$^\circ$-035$^\circ$, or equivalently 145$^\circ$-215$^\circ$, the remnant configuration is expected to be a quasiuniform state with no domain wall in the half-ring. Conversely, for the remaining field ranges, a domain wall is expected to be nucleated from the larger pad regions and propagate into the half ring until it is aligned with the field. In this case, the resistance of the structure is seen to decrease when a domain wall is present. However as confirmed by previous work, for this initial size regime the reduction can be attributed primarily to the AMR effect which, due to the component of magnetization within the wall that is directed off-axis to the wire, leads to a reduction of the wire resistance. For the second resistance state (495 $\Omega$) we still see the two previous levels corresponding to no domain wall ($A$) and a domain wall in the main part of the wire ($B$), however a third level is now evident for angles in the vicinity of the direction of the notch position (90/270$^\circ$). For a half ring of constant width, the domain wall is expected to be nucleated close to the position of the applied field due both to the curved geometry, as well as to the soft magnetic properties of the Permalloy. However, in the presence of significant constrictions the domain wall can be attracted to the resulting potential well from a large distance away. Hence this resistance level can be attributed to a new resistance state for a domain wall within the notch ($C$). As can be seen, with increasing resistance of the nanocontact, or equivalently for decreasing size of the contact at the notch, the resistance state corresponding to the DW in the notch in general becomes broader and larger. The change in resistance associated with the DW in the notch is plotted as a function of the resistance in Figure~\ref{etoile} (b). Here the resistance change, $\Delta R$, is calculated as the size of the peak corresponding to the DW in the notch, normalized with respect to the level with no DW in the structure \big($\Delta R = (C-B)/A$\big). As such for $\Delta R=0$ there is a difference in the contact resistance for a DW in the wire and in the notch, whereas for $\Delta R>1$ the resistance for a DW in the notch is larger than that with no DW in the structure. Strikingly it can be seen that for the narrowest contacts the resistance of state $C$, corresponding to a DW in the notch, even exceeds that of state A, and hence the effect can not be described solely based on a different size of the AMR for the particular spin state of the domain wall at the notch (which would reduce the resistance), but rather there must also exist an \emph{intrinsic} DWMR contribution which dominates for this size regime. While previous work has reported both positive and negative DWMR for Py, here we find a positive effect for our contact sizes.
 
\begin{figure}[tb!]
\centering
\includegraphics[width=0.8\textwidth,keepaspectratio=true]{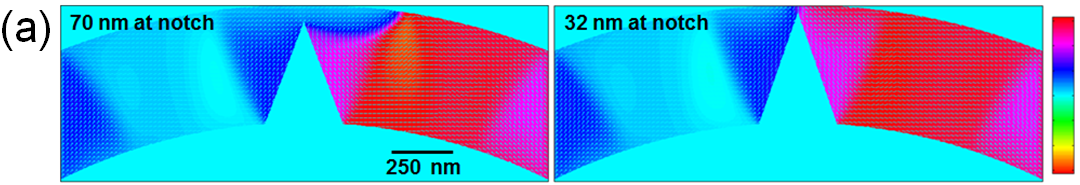}
\includegraphics[width=0.8\textwidth,keepaspectratio=true]{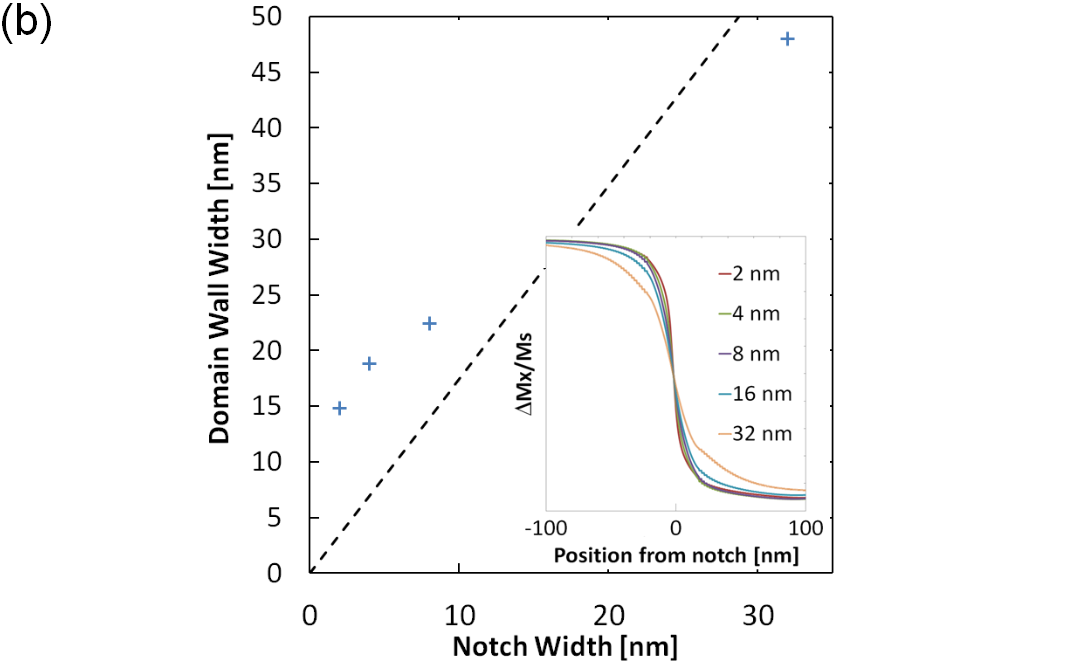}
\includegraphics[width=0.8\textwidth,keepaspectratio=true]{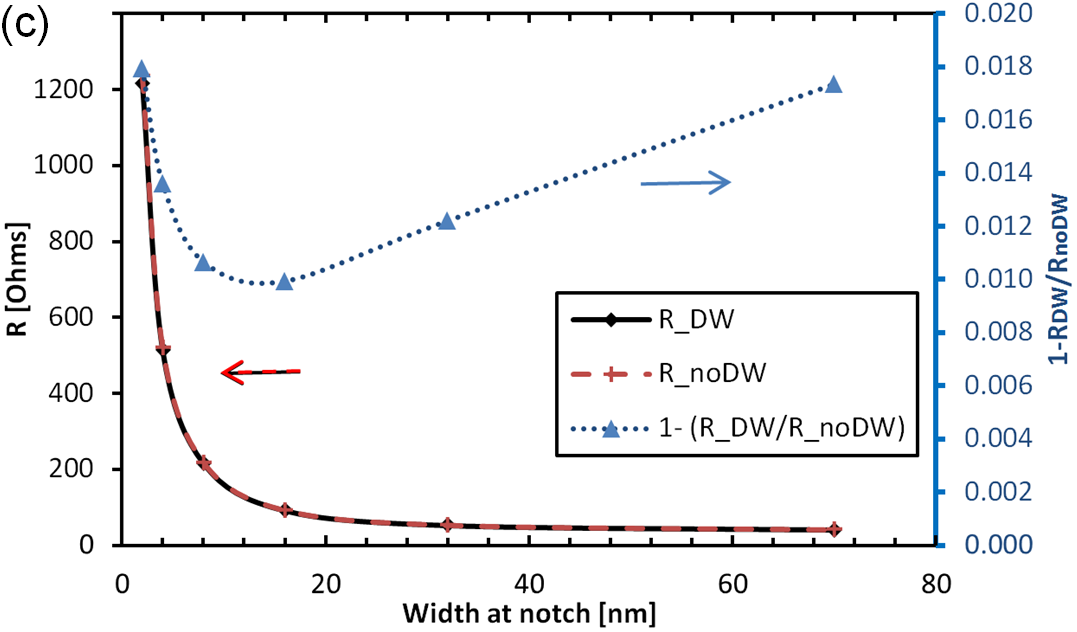}
\caption{(a) A series of micromagnetic simulations showing the evolution in DW spin structure as a function of contact size. (b) The evolution in DW width with the geometrical confinement, as extracted from the simulations. It is determined from the change in the x-magnetization profile due to the presence of the wall, as plotted in the inset. The dotted black line is a comparison to the previously determined experimental relation between DW width and wire width in Permalloy of $\delta \approx 1.74 \times w$~\cite{Laufthesis,headtohead}. (c) Evolution in AMR associated with the presence of a DW in the constriction as a function of the constriction size.}
\label{simulations}
\end{figure}

To gain insights into the evolution in the magnetic domain wall width with the nanocontact size, as well as the contribution from AMR, we perform multiscale micromagnetic simulations for the employed geometry. Figure~\ref{simulations} (a) presents typical simulations of the magnetization configurations as a function of the contact width, $w$. In all cases a transverse domain wall spin configuration is formed, as expected from the dimensions of the wires~\cite{pyphase} with a tilted transverse wall for the smaller notch evolving to a symmetric transverse wall for the larger notches, as previously observed in experiments~\cite{DWnotch}. Furthermore, as seen both in the images and in Figure~\ref{simulations} (b), the DW width, $\delta$, shows a monotonic dependence on the nanoconstriction size. This scaling of the wall size with the geometrical confinement is consistent with the predictions of Bruno~\cite{Bruno}, who showed that the increased exchange energy of the narrower wall is compensated by restricting the wall to a smaller volume, as has also been observed experimentally~\cite{SEMPADWsize, Laufthesis, headtohead, DWnotch}.  The AMR contribution from the changing DW is shown in Figure~\ref{simulations} (c). In such a notched system, the magnitude of the AMR component is an interplay between the confinement of the DW, which reduces the off-axis magnetization components and the confinement of the current, which means that the signal is increasingly dominated from the central region. In our case we find that in general the AMR contribution from the DW increases for smaller constrictions. However the size of the AMR contribution is rather small, on the order of 1-2\,\%, in agreement with previous reports of AMR in Py nanocontacts~\cite{AMRnanocontactsKlaui, BMRKlaui}, with sub 1\,\% changes for evolving contact size within the studied range, which is much smaller than the total effects we observe, and hence we now consider alternative explanations for the DWMR we observe.

\section{Discussion}
\label{sec:dis}
\begin{figure}[tb!]
\centering
\includegraphics[width=1.0\textwidth,keepaspectratio=true]{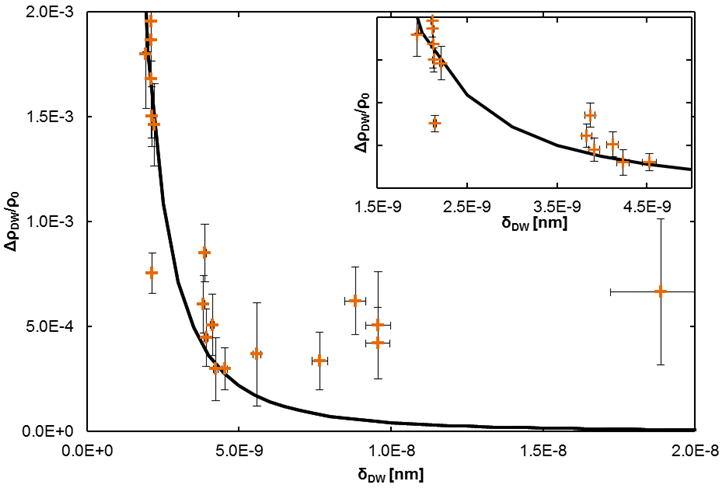}
\caption{Evolution of scaled domain wall resistivity with the calculated domain wall width. The solid line is a fit to Equation~\ref{fiteqn}. The inset shows an enlarged view of the data for small domain wall widths.}
\label{mainplot}
\end{figure}

In order to understand the data, it is instructive to fit the data to a power function of the following form, as predicted by theory~\cite{Viretmistracking, LevyZhang, banddDWMR, wireDWMR, bandsDWMR, bandsDWMR2}:

\begin{equation}
\frac{\rho_{DW}}{\rho_0}(\delta_{DW})=C \times (\delta_{DW})^\beta
\label{fiteqn}
\end{equation}
where $\rho_{DW}$ is the domain wall resistivity, $\rho_0$ is the material resistivity and $C$, $B$ are constants~\cite{Viretmistracking}. The resistivity of the domain wall can be defined based on the area of the contact, $A_c=\pi r^2$, and the domain wall width:

\begin{equation}
    \rho_{DW} = \Delta R_{DW} \times \frac{A_c}{\delta_{DW}}
\end{equation}
The contact radius, $r$, can be estimated from the contact resistance $R_c$ via Wexler's formula, which is an interpolation between the classical Maxwell expression of the conductance and the Sharvin expression valid in the ballistic regime~\cite{Wexler, AMRnanocontactsKlaui, NanoReeve}:
\begin{equation}
R_c=\frac{4}{3 \pi} \frac{\rho_0 l}{r^2}+ \gamma  \frac{\rho_0}{2r} \textrm{, with } \gamma=\frac{1+0.83(l/r)}{1+1.33(l/r)}
\label{wexler}
\end{equation}
$l$ is the electron mean-free path which is taken as 0.6\,nm~\cite{mfpref2} and the resistivity of Permalloy is taken as $\rho_0=2.95 \times 10^{-7} \Omega$m based on the relation in Ref.~\cite{mfpref}. The value of $R_c$ is calculated from the measured two-point resistances, corrected for the effective lead-resistance, which is determined to be $\sim429\pm5\,\Omega$ based on the initial contact dimensions~\cite{Loescherthesis}. The DW size is in turn calculated using the previously determined proportional scaling of the DW with wire width for rings ~\cite{Laufthesis,headtohead}, presented in Figure~\ref{simulations} (b), which has reasonable agreement with the simulations performed here for our notched system. The resulting data are presented in Figure~\ref{mainplot}. For larger contacts, Equation~\ref{wexler} returns a relatively large error in the contact size and correspondingly also the domain wall size, since there is some uncertainty in the original lead resistance. However as the contact is narrowed down, the increase in resistance is expected to originate solely from the notch region where the current density is highest and the electromigration therefore occurs and hence the junction region resistance increasingly dominates the overall two-point resistance.  

The line represents a fit to the data using Equation~\ref{fiteqn} which is weighted based on the strongly varying errors in the DW width~\cite{Loescherthesis}, yielding an exponent of $B=-2.31\pm0.39$. For the larger domain wall widths it can be seen that there is some deviation from the fit, with the data not tending to zero domain wall resistivity for the widest widths, indicative of additional contributions such as the AMR effect. However as confirmed by the micromagnetic simulations, the AMR effect is small for our geometry and does not dominate for the smaller constrictions. Furthermore, Yuan et. al predict the existence of an additional spin-orbit coupling induced contribution to the resistivity~\cite{SOCDWMR}, which would lead to a similar plateau for large domain widths and would also be consistent with the presented data. The observation of an intrinsic positive DWMR that scales with the contact size in Py is in contrast to our previous measurements for both larger and smaller contacts, since in the former case only AMR dominated signals were determined~\cite{AMRnanocontactsKlaui}, whereas in the latter both positive and negative DWMR was seen, the size and sign of which was primarily determined by the atomic co-ordination at the contact position~\cite{BMRKlaui}. The scaling of the domain wall resistivity with domain wall size has been theoretically studied by a number of approaches~\cite{MarrowsDWreview}, with certain models predicting a positive contribution to the domain wall resistivity which has an inverse scaling relation with the domain wall size ~\cite{Viretmistracking, LevyZhang, banddDWMR}. A semi-classical model by Viret et al. considered the ability of an electron to track the local magnetization direction as it traverses a domain wall~\cite{Viretmistracking}. As the width of the wall decreases, the non-adiabaticity of the majority electron spin alignment with the local wall spin direction increases. Hence changes to the dynamic non-adiabaticity parameter of the system are also expected, as hinted at by experimental depinning measurements. In the model, the electrons precess in the locally canted exchange field of the DW and depending on the rate of precession and the magnetization gradient they can track the changes to a greater or lesser extent. The resulting mistracking results in extra scattering and a corresponding increase in the DW resistivity which was found to scale with the square of the domain wall width. An equivalent scaling has been predicted by Levy and Zhang who treated the problem quantum mechanically~\cite{LevyZhang} and showed that the mixing of the spin channels is the origin of the enhanced resistivity, and other works also arrive at similar results~\cite{banddDWMR, wireDWMR}. In our case, the observed exponent of $B=-2.31\pm0.39$ agrees with these theories. Similar results have been seen for the PMA Pt/Co/Pt system, where Ga ion implantation was employed to modifiy the anisotropy, $K$, of the films and thereby set the DW width, which scales as $\pi \sqrt{A/K}$, where $A$ is the exchange stiffness~\cite{TuneDWMR}. In that work also a $1/\delta^2$ dependence was determined. Here, we reveal a comparable scaling dependence for an in-plane system consisting of a single layer and with DW width tailoring via geometrical confinement.

We now use our theoretical calculations to understand the regimes of validity of the different behaviours in more detail. We perform theoretical calculations of DWMR by solving a quantum-mechanical scattering problem, as described in section \ref{sec:DcConductivity}.

\begin{figure}[t]
\centering
(a)\includegraphics[width=0.8\textwidth]{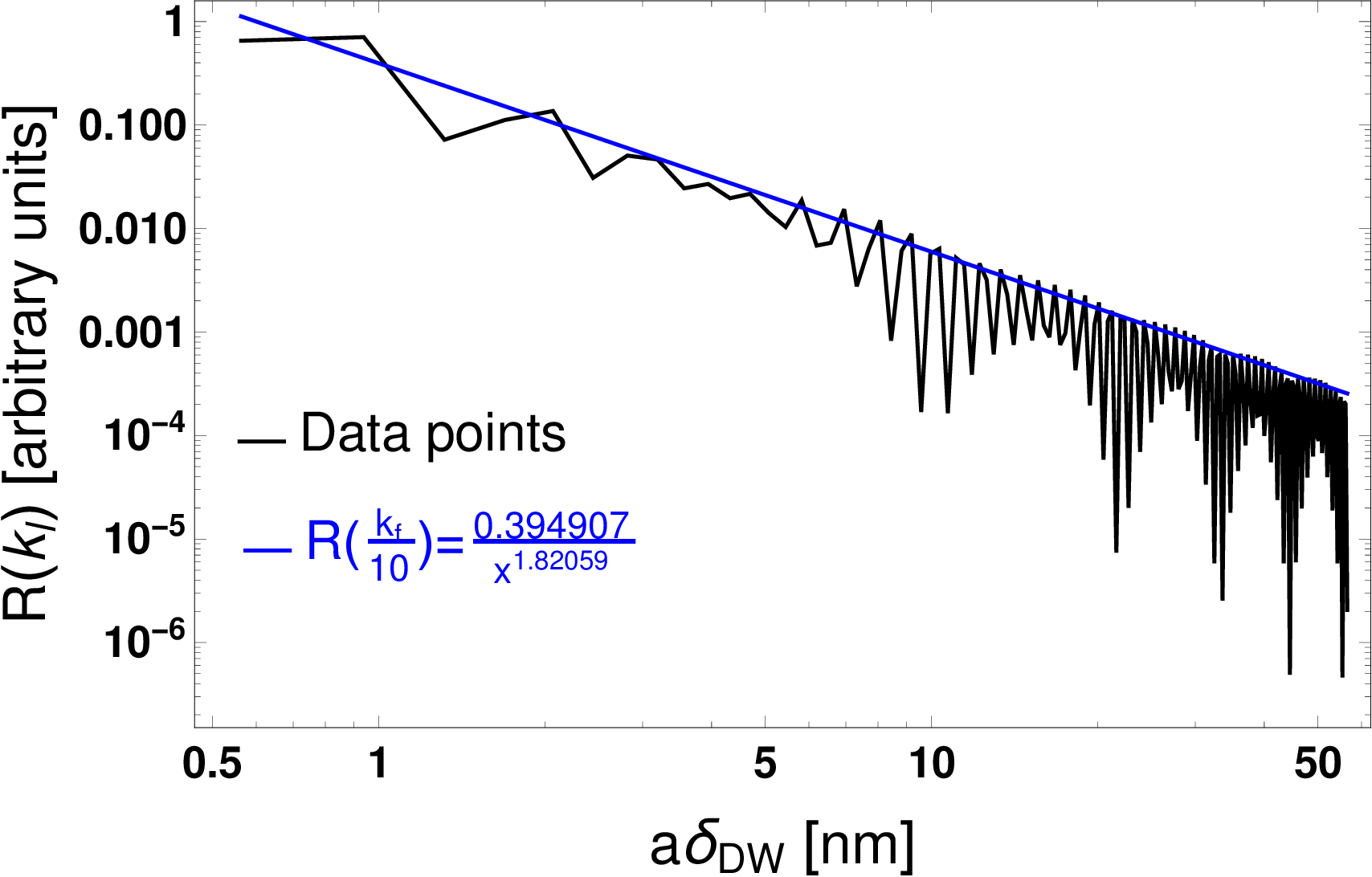}
(b)\includegraphics[width=0.8\textwidth]{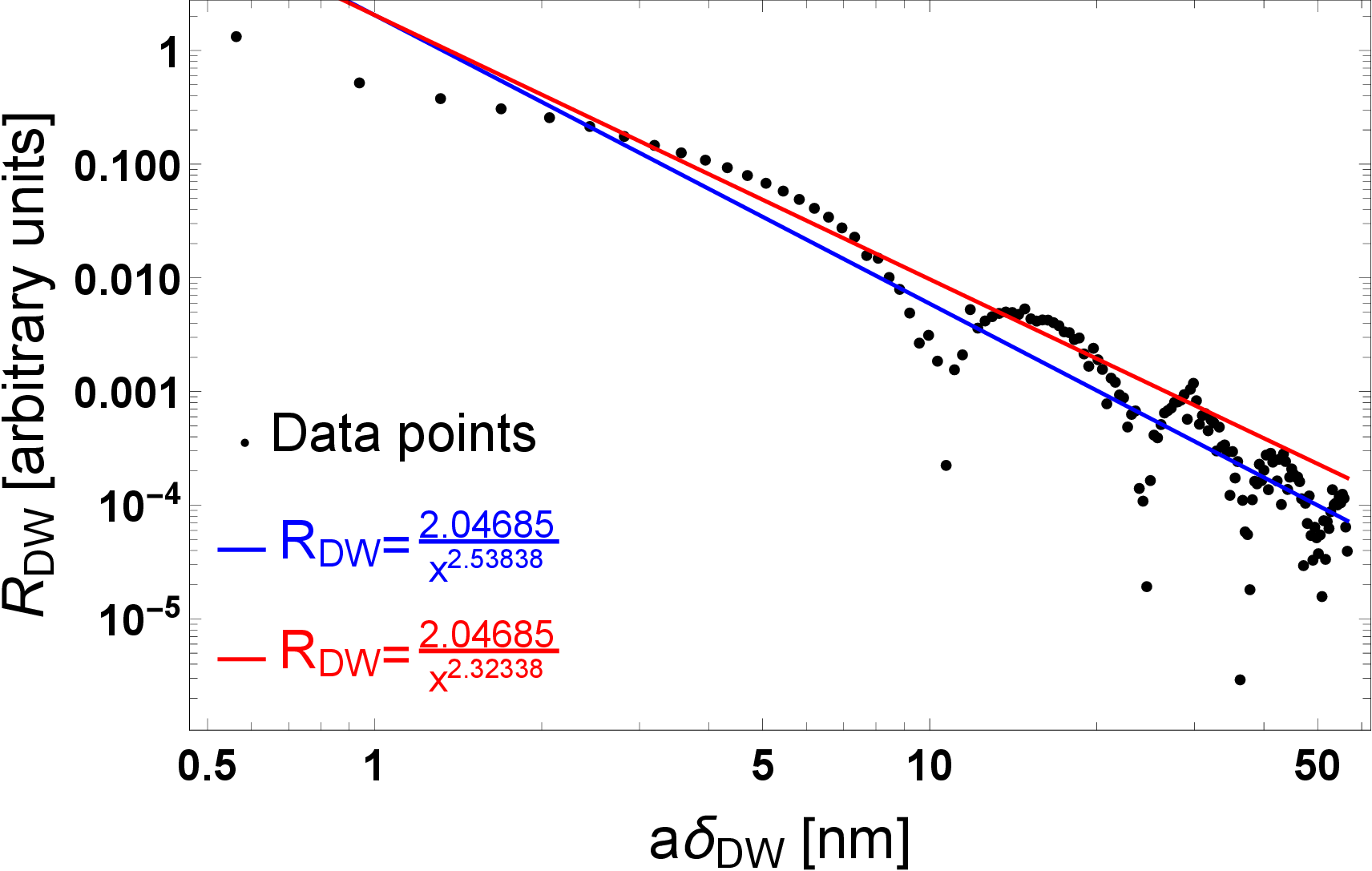}
\caption{\label{fig:DWMRvsL} (a) Calculated DWMR of a single channel of $k=\frac{k_f}{10}$ as a function of $\delta_{\rm DW}$. The maxima of the resistance for this single momentum scales with $\delta_{\rm DW}^{-1.82}$. For each value of the incident particle's momentum we obtain a different exponent and as we sum over all contributions up to Fermi momentum to get the total resistance the exponent diverges from inverse square as a result of oscillations. (b) DWMR ($R_{\mathrm{DW}}$) as a function of domain wall width ($\delta_{\mathrm{DW}}$) depicted on a log-log-plot. The DWMR fits to $1/\delta_{\mathrm{DW}}^{2.53}$ (blue line) but for all DW-widths, oscillations in the resistance occurs. The red line shows a fit to the maxima of the total DWMR and scales with $\delta_{\rm DW}^{-2.32}$ in excellent agreement with the experimentally measured exponent. Actual data shown by black dots.}
\end{figure}

It can be seen that for a single given incoming wave-number there are characteristic oscillations in the reflection and transmission coefficients due to the precession of the electron spins that may or may not be commensurate with the domain wall width. The contribution of a single channel to the DWMR is depicted for $k_{\uparrow}^{\rm l}= \hat{\bf z} \cdot \bm{k}_{\rm f}=\dfrac{k_{\rm f}}{10}$ in Fig.~\ref{fig:DWMRvsL}(a). The data for a single $k_{\uparrow}$ shows strong oscillations and a slow decay with $\delta_{\rm DW}$ which is changed by averaging over incoming directions in Eq. \ref{eq:totalDWMR} . The resulting calculated total DWMR is plotted in Fig.~\ref{fig:DWMRvsL}(b) in a log-log plot. The fit to the data shows that the average fit to the DWMR scales as $1/\delta_{\mathrm{DW}}^{2.53}$ (blue line), and the fit to the maxima of the DWMR scales with $1/\delta_{\mathrm{DW}}^{2.32}$ (red line) in excellent agreement with our experiments and earlier theoretical results~\cite{Tatara2000, BalDWMRscaling, bandsDWMR, bandsDWMR2}. This reveals that the observed scaling of the effects is robust and is preserved on transition from the diffusive~\cite{Viretmistracking} to the ballistic regime as studied here and hence could be usefully employed in a device setting.

\section{Conclusion}%
\label{sec:con}
In conclusion, we have studied the evolution in domain wall resistivity with the size of the domain wall in electromigrated permalloy nanocontacts. By employing a robust technique to create clean, magnetostriction-free nanowires with tailored size and with domain walls stable at remanence, we are able to demonstrate for narrow domain walls in narrow constrictions a clear positive intrinsic contribution to the domain wall resistivity which is not due to the anisotropic magnetoresistance effect. This finding is supported by micromagnetic simulations, which additionally demonstrate the geometrical confinement of the domain wall as the nanocontact is reduced in size by electromigration. The domain wall resistivity is found to scale with the domain wall width as $1/\delta_{DW}^{2.31 \pm 0.39}$, in excellent agreement with transport calculations which yield a dependence of $1/\delta_{DW}^{2.32}$, confirming increased scattering of the electrons as they traverse narrower domain walls which provide a more abrupt transition in the magnetization and hence a sharper potential step. While large domain wall magnetoresistance effects have previously been reported, they were often not particularly reliable or depended very delicately on the contact configuration. Hence the robust demonstration of a regime where large effects are seen and which scale monotonically with the geometrical confinement is of particular interest for proposed devices where notches are routinely employed as geometrical pinning centres for domain walls and where large magnetoresistance effects are beneficial.  

\section*{Acknowledgments}

Funded by the Deutsche Forschungsgemeinschaft (DFG, German Research Foundation) - project number 268565370/TRR173 through the collaborative research centre SFB/TRR 173 Spin+X, "Spin in its collective environment" (Projects B02, A03 and A10) as well as the Graduate School of Excellence Materials Science in Mainz (GSC266). This work was supported in Rzesz\'ow University of Technology by the National Science Center in Poland as research Project No.~UMO-2017/27/B/ST3/02881.

%


\end{document}